\documentclass[10pt,a4paper,twocolumn]{article}
\newcommand{\surveyshorttitle}{Prompt Corpora and GEO Visibility}
\newcommand{\surveydate}{Version dated 6 September 2026}
\usepackage[left=16.5mm,right=16.5mm,top=18mm,bottom=18mm,headsep=6mm,footskip=10mm]{geometry}
\usepackage{fontspec}
\usepackage{amsmath,amssymb,booktabs,array}
\usepackage{microtype}
\usepackage{color}
\definecolor{SurveyNavy}{rgb}{0.08235,0.22745,0.35686}
\definecolor{SurveyTeal}{rgb}{0.08627,0.49020,0.49804}
\definecolor{SurveyGrey}{rgb}{0.36863,0.40784,0.44706}
\definecolor{SurveyPale}{rgb}{0.92941,0.95686,0.96471}
\usepackage[authoryear,square]{natbib}

\usepackage{xurl}
\usepackage[colorlinks=true,linkcolor=SurveyNavy,citecolor=SurveyTeal,urlcolor=SurveyNavy]{hyperref}
\usepackage{fancyhdr}
\makeatletter
\renewcommand\section{\@startsection{section}{1}{\z@}{-2.2ex \@plus -0.5ex \@minus -0.2ex}{0.8ex \@plus 0.2ex}{\normalfont\fontsize{12}{14}\selectfont\bfseries\color{SurveyNavy}}}
\renewcommand\subsection{\@startsection{subsection}{2}{\z@}{-1.7ex \@plus -0.4ex \@minus -0.2ex}{0.5ex \@plus 0.1ex}{\normalfont\normalsize\bfseries\color{SurveyTeal}}}
\renewcommand\tableofcontents{\section*{\contentsname}\@starttoc{toc}}
\makeatother
\newcommand{\surveyfrontmatter}[6]{%
  \twocolumn[{%
    \begin{minipage}{\textwidth}
    \begin{center}
      {\small\bfseries\color{SurveyTeal}#1\par}
      \vspace{1.5em}
      {\fontsize{17.28}{20.5}\selectfont\bfseries\color{SurveyNavy}#2\par}
      \vspace{0.45em}
      {\fontsize{12}{14.4}\selectfont #3\par}
      \vspace{1.2em}
      {\normalsize #4\par}
      {\footnotesize\href{mailto:#5}{#5}\par}
      \vspace{1.1em}
      \setlength{\fboxsep}{4pt}%
      \colorbox{SurveyPale}{\parbox{0.92\textwidth}{\footnotesize #6}}
    \end{center}
    \vspace{1.0em}
    \end{minipage}%
  }]
}

\usepackage[english]{babel}
\hypersetup{pdftitle={Measuring GEO Visibility: Prompt Corpora Define the Answer Market},pdfauthor={Olivier Martinez},pdfsubject={GEO visibility measurement: validity, partial identification, and source attribution},pdfkeywords={GEO, generative engine optimization, visibility measurement, prompt corpora, answer market, source citations}}

\begin{document}
\selectlanguage{english}
\surveyfrontmatter{CRITICAL LITERATURE SURVEY}{Measuring GEO Visibility: Prompt Corpora Define the Answer Market}{A Critical Survey of Validity, Partial Identification, and Source Attribution}{Olivier Martinez}{olivier.martinez2@sciencespo.fr}{\textbf{Version note.} Critical survey and methodological analysis; revised 6 September 2026. The demonstrator recalculates published aggregates; no new experimental data collection is reported.}
\section*{Abstract}
\begingroup\clubpenalty=10000 To measure the visibility of a source or brand in a generative engine’s answers, a set of prompts is submitted to the engine, and displayed sources, citations, or mentions are recorded according to specified rules. These observations are aggregated into a GEO (\emph{generative engine optimization}) visibility score. The choice of prompts determines which situations are evaluated; the weights determine their relative importance in the calculation. We call this weighted set of answer opportunities an “answer market.” This constructed market does not necessarily represent actual user demand.

Prompt wording can alter document retrieval, the sources brought into competition, and the answers produced. Once these answers have been generated, calculating the visibility score requires identifying the source appearances, citations, or mentions of interest. If this task is assigned to a language model, a second instruction enters the process: the instruction telling the model how to evaluate the answers. This instruction can change the scoring even when the evaluated answers remain identical. Our critical survey examines how these choices help define what a GEO score measures. It draws on research that examines whether the indicators used measure the intended phenomenon, as well as research on total survey error and information retrieval evaluation. The proposed framework makes explicit the evaluated situations and their annotation, prompt formulations, execution conditions, weights, and scoring rules.

Two calculations illustrate the consequences of these choices. Using published aggregate rates from three selected subcorpora in Grossman et al., two weighting schemes yield generative-answer activation rates of 39.7\% and 70.5\%, without changing the rates within each subcorpus. These results concern whether a generative answer appears; they do not measure brand visibility. A fictional example shows that reweighting can reverse the ranking of two systems when each performs better in different categories of situations.

When weights are unknown or remain to be chosen, the framework proposes reporting a set of admissible scores. It distinguishes two approaches: determining the values compatible with the data and assumptions about a target population, which concerns partial identification; and examining how the score varies across weighting conventions, which concerns normative sensitivity.

The presence of a citation is insufficient to establish a source’s contribution to an answer. To examine this contribution, the article defines a comparison between answers produced with and without a source in a controlled documentary context. It distinguishes this contrast from an intervention on the full engine in the presence of competing sources. Together, these contributions provide a framework for interpreting GEO scores, supported by reproducible calculations. No new experiments are reported; the general empirical validity of the proposed protocol remains to be assessed.\par\endgroup
\par\begingroup\interlinepenalty=10000\noindent \textbf{Keywords:} generative engine optimization (GEO); GEO visibility measurement; prompt corpora; answer market; source citations; construct validity; partial identification; source attribution.\par\endgroup
\begingroup\footnotesize\setlength{\parskip}{0pt}\tableofcontents\endgroup
\section{Introduction}
Our focus is visibility measurement for generative engine optimization (GEO). A prompt corpus defines an \textbf{answer market}: the weighted answer opportunities over which displayed source presence and citations are observed. A \textbf{GEO visibility score} aggregates those observations using a declared detection rule and denominator; brand mentions require their own rule.

This constructed market need not represent human demand. Stating that a source is cited in 40\% of answers is therefore insufficient: are these answers to open-ended requests, prompts naming the source, or comparison requests? Which engine, period, and weights? These choices define the scope of the score.

Interpreting this score requires distinguishing two levels. A prompt can shape the response produced; the choice of corpus, weights, and scoring rules helps define the quantity to which the score refers. An influence on the output is a behavioral fact; participation in the definition of measurement is a property of the evaluation design.

In generative systems, these two levels interact: a prompt can perform three coupled roles. It selects and realizes a situation (an annotated evaluation case, rather than an assumed observable latent information need; formalized in Section 4.1); it acts on the procedure producing the answer, including search activation, rewritten queries, and retrieved documents; finally, when scoring relies on an LLM judge, a second instruction helps construct the score. The document set can thus become endogenous to the intervention, and the same output can be evaluated differently depending on the judging instrument. The problem therefore extends beyond the ordinary dependence of an average on its sample. These properties are neither universal across benchmarks nor without precedents; their combination makes merely reporting “prompt sensitivity” insufficient.

Our thesis is that the corpus, its formulation policies, and the scoring apparatus constitute an \textbf{active measurement instrument}: some choices define the target, others elicit system behavior, and still others transform outputs into observations. Distinguishing them makes it possible to decide what should be fixed, estimated, or subjected to sensitivity analysis. The system's properties exist independently of the corpus; this does not justify presenting a score without specifying the instrument through which they are observed.

Critiques of benchmark generality and validity did not begin with generative engines. Raji et al. challenge the move from bounded performance results to general claims about AI; Bowman and Dahl examine shortcomings in NLU benchmark design and interpretation \citep{raji2021everything,bowman2021benchmarking}. Multi-prompt evaluations and request-conditional rankings make this problem concrete for language models \citep{mizrahi2024multiprompt,frick2025p2l}. This article connects these critiques to a computable object and precisely defined source interventions.

We propose four contributions of different scope: a taxonomy of the instrument's functions; sets of scores compatible with explicit constraints on weights; a source-removal contrast distinguishing conditional attribution from effects in production; and an arithmetic demonstrator using published aggregates. The identities and examples establish consequences under stated assumptions. The comparative empirical usefulness of the documentation protocol remains to be evaluated.

\section{Scope, Selection, and Traceability of the Survey}
This critical literature survey is current as of 31 August 2026. It starts from a dossier on prompt corpora and checks the selected references against public primary sources. The revision extends this corpus to benchmark validity, incomplete relevance judgments, total survey error, partial identification, item response theory, and citation attribution. Internal documents and preparatory exchanges support discovery; they do not constitute public scientific evidence.

Selection combines exact titles, keyword families, and backward citation tracing: \emph{prompt sensitivity}, \emph{measurement validity}, \emph{query sampling}, \emph{generative engine optimization}, \emph{incomplete relevance judgments}, \emph{total survey error}, \emph{partial identification}, \emph{item response theory}, and \emph{citation faithfulness}. Sources consulted include ACL Anthology, ICLR and NeurIPS proceedings, PMLR, arXiv, scholarly publishers, and institutional or author repositories. A reference is included if it informs at least one decision concerning the definition, selection, formulation, execution, scoring, or transport of measurement. Studies are not selected according to the direction of their results.

The corpus comprises forty-four references. For each, the accompanying register distinguishes publication status, version, primary URL, passages actually consulted, and use in the argument. It does not equate a bibliographic record with reading the full text. \textbf{Messick (1994) and Frick et al. (2025) are used at the level of their records and abstracts}: the former solely for the intellectual lineage of validity, the latter for the stated objective of prompt-conditional ranking. No detailed experimental result is inferred from them here. For Zobel, only the introduction was consulted; for Manski, the opening pages of the introduction. The demonstrator uses the benchmark construction in Section 3, the query counts in Table 1, and the activation rates in Section 4.1 of Grossman et al. \citep{messick1994validity,frick2025p2l,zobel1998reliable,manski2003partial,grossman2026disrupts}.

This selection is neither exhaustive nor independently coded twice. Documentary inclusion probabilities and the number of relevant studies not retrieved are unknown. We therefore calculate neither a frequency of findings in “the entire literature” nor an average of heterogeneous effects. The appendix applies our own corpus card to this selection: the aim is to make its construction inspectable, not to turn it artificially into a representative sample. Aggregate recalculations are reproducible; the articles' experiments are not reproduced.

The main text of Tamer's survey was read in full (abstract and Sections 1–5, pp. 167–192) in the primary author-hosted Review in Advance version, using paginated text extraction; the references were also consulted. This is not a claim to have independently verified the cited theorems or visually audited every formula. It provides the operative anchor for separating identification, estimation, and inference in Section 4.4 \citep{tamer2010partial}. For construct validity, Jacobs and Wallach's Section 3 was read in full; the register records the targeted scope of the remaining reading.

\section{From Construct to Active Instrument}
\subsection{Four Distinct Objects}
The \textbf{construct} is the phenomenon of interest, such as a capability, exposure, or attribution quality. Its \textbf{operationalization} translates it into observable situations and criteria. The \textbf{estimand} specifies the target quantity, including its unit and denominator. The \textbf{estimator} is the calculation rule applied to the data. Our methodological anchor for validity is Jacobs and Wallach's account of measurement in computational systems, including content validity and the possibility that several proxies share the same defect \citep{jacobs2021measurement}. Messick is retained as an intellectual antecedent, on the limited reading declared in Section 2. A string match can establish an exact mention; by itself, it establishes neither a favorable recommendation nor an informational contribution.

This distinction avoids a central ambiguity. If an external population has been defined in advance, changing the sample changes the estimate and its possible bias, not automatically the estimand. If the panel and its weights define the conventional population under study, changing those weights changes the target. A formulation change can reveal measurement error at a constant construct or genuinely change the task. These possibilities must be distinguished before any comparison.

Total survey error provides an integrated vocabulary for representation and measurement problems, including weights and data processing \citep{groves2010totalsurvey}. Wording effects studied in human questionnaires also provide a precedent, without assuming identical cognitive mechanisms in humans and models \citep{schwarz1999selfreports}. This survey adapts these distinctions to an instrument that can act on retrieval and can itself be scored by another model.

\subsection{Five Functions of Prompts in a Protocol}
A prompt can specify the task, select a situation, realize a need linguistically, condition the interaction, and impose an evaluable output form. Table 1 specifies the corresponding decisions before we examine their dependencies.

\begin{table*}[t]
\centering
\caption{Functions of prompts within the measurement protocol.}
\small
\renewcommand{\arraystretch}{1.12}
\begin{tabular}{@{}>{\raggedright\arraybackslash}p{\dimexpr 0.24\linewidth-1.3333333333333333\tabcolsep\relax}>{\raggedright\arraybackslash}p{\dimexpr 0.36\linewidth-1.3333333333333333\tabcolsep\relax}>{\raggedright\arraybackslash}p{\dimexpr 0.4\linewidth-1.3333333333333333\tabcolsep\relax}@{}}
\toprule
\textbf{Function} & \textbf{Embedded decision} & \textbf{Consequence for interpretation} \\ \midrule
Specify the task & Definition, choice, justification, or list & Elicited behavior and success criterion \\
Select situations & Included needs, languages, people, and domains & Support over which the score is informative \\
Realize the need & Words, format, examples, and precision & Sensitivity to formulation \\
Condition the interaction & History, candidates, and assumed knowledge & Available information and trajectory \\
Make the output evaluable & Required sources, length, or format & Produced answer and ease of scoring \\
\bottomrule
\end{tabular}
\end{table*}
These functions are not five independent variables. Requesting sources can change retrieval, the answer, and the citation denominator simultaneously. Translation can preserve intent while changing the accessible resources. Weights and the brand detector are not contained in the prompt text, but belong to the instrument. The taxonomy locates these decisions; it does not claim to decompose causal effects that have already been identified.

\subsection{Three Aims and a Precedent in Information Retrieval}
A \textbf{controlled benchmark} targets comparison on fixed cases; a \textbf{strategic panel} weights situations according to a normative priority; a \textbf{usage estimate} targets frequencies in a defined population, period, and interface. The first two are not deficient estimates of the third. They become misleading when presented as such.

Information retrieval test collections already combine documents, topics, and judgments \citep{voorhees2002philosophy,thomas2026validity}. \emph{Pooling} reveals a crucial precedent: documents to be judged are selected from systems' results, and incomplete judgments can affect measurement. Zobel raises this problem while reporting relative robustness in some comparisons; Buckley and Voorhees examine measures suited to incompleteness \citep{zobel1998reliable,buckley2004incomplete}. This proves neither that every pool biases every ranking nor that a citation is equivalent to a relevance judgment. The parallel concerns selection into observability. In a generative engine, this selection can additionally change under the influence of the prompt being evaluated.

\par\ifdim\dimexpr\pagegoal-\pagetotal\relax<8\baselineskip\relax\newpage\fi
\section{Defining the Target, Then the Compatible Values}
\subsection{Annotation and the System Boundary}
Let $X$ be a documentable raw unit: a usage event, a query with context, or a constructed vignette. Its target distribution is $R$. An annotation procedure $A_\eta(\cdot\mid X)$ produces an operational situation $U$, with guidelines, categories, and adjudication rules versioned by $\eta$. Its distribution is defined by $Q_\eta(B)=\int A_\eta(B\mid x)\,dR(x)$. Deterministic annotation is a special case. $U$ is therefore not presented as a latent “true need” that is spontaneously observable; identifying such a need would require an additional measurement model and validation.

The prompt $P$ follows a policy $K(\cdot\mid U)$; execution conditions $C$ follow $H(\cdot\mid U,P)$; the system $S$ produces $Y$. A descriptive protocol is the tuple $\mathcal T=(R,A_\eta,K,H,g)$, where $g$ transforms the output into a score. For a mean per unit:

\begin{equation}
\begin{aligned} \theta_S(\mathcal T)&=\mathbb E_{U,P,C}[m_S(U,P,C)],\\ m_S(u,p,c)&=\mathbb E_{Y\sim S(\cdot\mid u,p,c)}[g(Y,u,p,c)]. \end{aligned}
\end{equation}
The distributions in the expectation are $Q_\eta,K,H$. Conditioning on $U$ describes the evaluation; it does not grant the system privileged access to the annotation or a hidden need. Information actually transmitted must appear in its inputs.

The boundary between $S$ and $H$ follows a rule tied to the intervention under study. Prior inputs and settings that the design fixes or distributes belong to the protocol $H$; executed mechanisms and the variables they produce downstream belong to $S$. Thus, if the intervention concerns the prompt of a complete engine, rewritten queries and retrieved documents are mediators, not fixed conditions. Imposing a documentary context constitutes another protocol, concerning the conditional generator. $H$ can be averaged over or reduced to a fixed value; it does not mean “what the analyst does not average over.”

If $g$ is an LLM judge, its model, rubric, instruction, and the distribution $\mathcal L$ of its variants must also be declared. The judge's mean score can be written as $g_{\mathcal L}(y,u,p,c)=\mathbb E_{\Xi\sim\mathcal L}[\psi(y,u,p,c;\Xi)]$. Varying $K$ changes the interaction; varying $\mathcal L$ changes the scoring instrument. A crossed design can examine them separately, including on the same saved outputs. Annotation and the judge add identifiable choices to the protocol; documenting them does not suffice to guarantee their validity.

Finally, not all metrics are means of elementary scores. A brand's global share of citations is $\rho_S=\mathbb E[N_b]/\mathbb E[N_{\mathrm{tot}}]$ when the denominator is positive. It differs from the mean of per-answer ratios, which requires a convention for answers without citations. A probability of at least one exposure per user, in turn, requires a trajectory-level unit. These targets are not interchangeable.

\subsection{Origins of Weights and the Finite Estimator}
For $n$ situations and $k_i$ variants of each, let $w_i$ and $v_{ij}$ be normalized nonnegative weights. With $r_{ij}$ repetitions, the estimator is:

\begin{equation}
\begin{aligned} \widehat\theta_S&=\sum_i w_i\sum_j v_{ij}\overline Z_{ij},\\ \overline Z_{ij}&=\frac{1}{r_{ij}}\sum_{\ell=1}^{r_{ij}}g(Y_{ij\ell},u_i,p_{ij},c_{ij\ell}),\\ \sum_i w_i&=1,\qquad\sum_jv_{ij}=1. \end{aligned}
\end{equation}
Three origins of weights must be distinguished: inclusion probabilities and calibration in a survey; documented frequencies in logs with their coverage; and normative priorities defining a panel. Uniform weights on available formulations define a variant policy; they do not automatically estimate how humans formulate a need. Estimated weights must retain their own uncertainty. Unknown frequencies do not become known because a simple average is convenient.

The hierarchical structure prevents a family from gaining weight simply by multiplying its paraphrases. Conversely, averaging all rows in a file implicitly weights families by their size. Deduplicating and then uniformly weighting unique texts changes a distribution of events into a distribution of types. These operations are admissible if their target is declared.

\subsection{Ranking Reversal: A Conditional Possibility}
Consider two fictional systems and the probabilities stipulated in Table 2. No experimental result is represented.

\begin{table}[t]
\centering
\caption{Fictional analytical example: success probabilities by stratum.}
\small
\renewcommand{\arraystretch}{1.12}
\begin{tabular}{@{}>{\raggedright\arraybackslash}p{\dimexpr 0.5\linewidth-1.3333333333333333\tabcolsep\relax}>{\raggedright\arraybackslash}p{\dimexpr 0.25\linewidth-1.3333333333333333\tabcolsep\relax}>{\raggedright\arraybackslash}p{\dimexpr 0.25\linewidth-1.3333333333333333\tabcolsep\relax}@{}}
\toprule
\textbf{Stratum} & \textbf{System A} & \textbf{System B} \\ \midrule
Open-ended discovery & 0.30 & 0.50 \\
Named verification & 0.90 & 0.60 \\
\bottomrule
\end{tabular}
\end{table}
With $\lambda$ as the weight on discovery, $\theta_A(\lambda)=0.90-0.60\lambda$ and $\theta_B(\lambda)=0.60-0.10\lambda$. Their difference is:

\begin{equation}
\theta_A(\lambda)-\theta_B(\lambda)=0.30-0.50\lambda.
\end{equation}
At $\lambda=0.2$, the scores are $0.78$ and $0.58$; at $\lambda=0.8$, they are $0.42$ and $0.52$. The ordering reverses at the threshold $0.60$ because the profiles cross between strata. If A dominates B in every stratum, common nonnegative weights cannot reverse the ordering. The example establishes this conditional possibility, not universal ranking instability.

\subsection{Admissible Sets and Partial Identification}
When situation or variant weights have not been settled, consider a class $\Pi$ of joint masses $\pi_{ij}=w_iv_{ij}$, with $\pi_{ij}\geq0$ and $\sum_{ij}\pi_{ij}=1$. Known margins, frequency intervals, or constraints on variants can restrict this class. For conditional means $\mu_{ij}$, distinguish the target set from its plug-in estimate:

\begin{equation}
\begin{aligned} \Theta(\Pi)&=\left\{\sum_{ij}\pi_{ij}\mu_{ij}:\pi\in\Pi\right\},\\ \widehat{\Theta}(\Pi)&=\left\{\sum_{ij}\pi_{ij}\overline Z_{ij}:\pi\in\Pi\right\},\\ \widehat\theta_-&=\min\widehat{\Theta}(\Pi),\\ \widehat\theta_+&=\max\widehat{\Theta}(\Pi). \end{aligned}
\end{equation}
For a nonempty compact convex class, the linear image is an interval and the extrema are attained; linear constraints yield two linear programs. A finite list of scenarios instead yields a finite set, whose envelope can be reported. The class must be justified before inspecting the results; otherwise, it provides another way to select a favorable ranking.

Two interpretations must not be confused. If the weights define several legitimate strategic priorities, $\Theta$ is a \textbf{sensitivity envelope across estimands}. If the conditional means $\mu_{ij}$ are identified on the support and $H,g$ are fixed, an empirical target distribution may remain unknown. When $\Pi$ describes exactly the masses compatible with the observations and assumptions, $\Theta$ is then an \textbf{identified set}. A class chosen for convenience does not deserve the latter designation. Following Tamer, identification concerns what the observation distribution and assumptions determine, while estimation and inference introduce sampling uncertainty \citep{tamer2010partial}. Uncertain annotation can similarly be addressed through a declared class of procedures $A_\eta$.

In the fictional example, imposing $\lambda\in[0.2;0.8]$ gives $\Theta_A=[0.42;0.78]$ and $\Theta_B=[0.52;0.58]$. The difference under \textbf{the same weight} belongs to $[-0.10;0.20]$. Freely subtracting the two intervals would yield $[-0.16;0.26]$, an envelope too wide for this coupled comparison. A is strictly better for any class whose maximum weight is below $0.60$; B is strictly better if the minimum weight exceeds that threshold. A class crossing the threshold permits both orderings. This check supplies an operational robustness criterion.

The set $\widehat{\Theta}(\Pi)$ and its extrema estimate their population counterparts; they are not confidence intervals. If simultaneous intervals $[l_{ij},u_{ij}]$ cover the means of all cells at level $1-\alpha$, an outer envelope is $[\min_{\pi\in\Pi}\sum\pi_{ij}l_{ij},\max_{\pi\in\Pi}\sum\pi_{ij}u_{ij}]$. This outer envelope separates sampling uncertainty from weighting choices and can be conservative: simultaneous marginal bounds need not exploit the joint constraints on the means. Coverage of the whole set here is conditional on a fixed class $\Pi$; estimated constraints require propagating their own uncertainty. An empty class calls for diagnosing the restrictions, not reporting extrema. An unobserved cell with a score in $[0,1]$ retains these bounds in the absence of an assumption; assigning zero to it does not correct a coverage failure.

\subsection{Transport and Temporal Comparison}
Absorbing formulations and conditions into $m_t(u)$ yields the identity:

\begin{equation}
\begin{aligned} \theta_1-\theta_0&=\int[m_1(u)-m_0(u)]\,dQ_0(u)\\ &\quad+\int m_1(u)\,d[Q_1-Q_0](u). \end{aligned}
\end{equation}
The first term compares mean responses on the initial population; the second varies composition at final responses. The reference ordering is a descriptive choice, not a unique causal attribution. A change in variant policy, environment, or judge enters $m_t$. Changing annotation can alter $Q_\eta$ and the conditional means: the decomposition then requires a common situation space or an explicit correspondence between coding schemes. Its terms require observations or assumptions concerning cells crossed between periods.

For $0\leq m\leq1$, the bound $|\theta(Q)-\theta(Q')|\leq\mathrm{TV}(Q,Q')$ uses the convention $\mathrm{TV}(Q,Q')=\sup_B|Q(B)-Q'(B)|$, or half the $L^1$ distance when densities exist. It supplies no automatic correction. If $Q$ is constrained only to a total variation ball around a documented distribution, it gives an outer envelope; exploiting the profiles across strata can tighten it.

\section{What the Literature Shows about Formulations}
The following studies primarily evaluate capabilities on language-model benchmarks, rather than source visibility. They establish that formulation policies can materially affect outputs and their scoring; this makes $K$ a substantive choice in a GEO measurement protocol, not a parameter to omit. Their effect sizes do not transfer automatically to citation or mention rates. The connection is methodological: distinguish equivalent variants, changed needs, and changes in scoring, then test those distinctions on the visibility outcome of interest. Section 6 extends this argument to the answer market's unit. A corpus of initial prompts measures first-answer opportunities; a corpus of trajectories also includes opportunities created by clarification and later turns. Multi-turn research motivates treating these as different targets, without supplying the distribution of real purchasing journeys.

\subsection{Genuine but Conditional Empirical Sensitivity}
Sclar et al. study format changes designed to preserve the task while keeping example selection and order fixed. On the models and tasks considered, these changes can produce substantial differences and alter comparisons. However, the grammar of variants defines the tested space; it is not a distribution of human formats \citep{sclar2024formatspread}. Mizrahi et al. extend evaluation to multiple instruction paraphrases and show the value of distinguishing mean performance from the best performance obtained. Their design covers 20 models and 39 tasks, with 6.5 million evaluation instances; this scale does not turn synthetic paraphrases into a sample of usage \citep{mizrahi2024multiprompt}.

PromptEval proposes efficiently estimating a distribution of scores over prompts rather than reducing evaluation to one formulation \citep{polo2024prompteval}. ReliableEval examines stochastic evaluation and moment estimation; perturbations do not concern words alone, since some configurations also vary evaluated examples or demonstrations \citep{lior2025reliableeval}. These studies provide methods for a defined distribution of variants. They do not, by themselves, provide the target distribution of human requests.

The common lesson is therefore more precise than a universal requirement for “more prompts.” Researchers must specify which variations belong to the target population, which test an invariance, and which seek maximum performance. Optimizing an instruction and then reporting its score measures a system accompanied by an adaptation procedure. Comparing that score with one from another system evaluated without a comparable budget can confound system capabilities with the effort devoted to adaptation. Selection data and the final test must also be separated: choosing the best instruction on the same observations used to report its performance creates selection optimism. A comparable budget does not correct this reuse of the test.

\subsection{Scoring Can Produce Part of the Sensitivity}
Hua et al. show that some variations attributed to prompts also depend on scoring methods. Accepting semantically correct answers can reduce the dispersion observed in the benchmarks studied \citep{hua2025artifact}. This does not mean that all recommendation variations are artifacts or that an LLM judge is an error-free reference. A formulation change can alter the format of a correct answer, its content, or both.

The implication is to validate the observed answer together with its transformation into a score. A brand detector can confuse a name with a common word; a parser can miss a recommendation expressed without a list; an evaluator can treat a negative mention as an endorsement. When $g$ relies on an LLM, its own instructions form a second set of prompts. Evaluation then has at least two layers sensitive to formulation: the system under study and the scoring instrument.

\subsection{Not Every Response Variation Is a Defect}
A reformulation is equivalent relative to a task and a judging rule, not merely by linguistic similarity. Adding “for a small nonprofit” or “available in France” can legitimately change a recommendation. Requiring an identical list in such cases would reward insensitivity to constraints. CheckList accordingly distinguishes transformations under which invariance is expected from those under which a directional change is desirable \citep{ribeiro2020checklist}.

In a preprint on commercial recommendations, Jack et al. distinguish cosmetic rewrites from variants that add constraints and compare them with identical reruns. Recommendation similarity is lower across variants in their design. However, the control families and configurations are not entirely identical; the enriched variants also alter the need. The defensible reading concerns the robustness of these recommendations under these transformations, not a general impossibility of measuring presence \citep{jack2026brittleness}.

\section{Building a Corpus: Provenance, Coverage, and Interaction}
\subsection{Real, Plausible, and Representative}
A prompt's human origin does not establish its representativeness. WildChat collects conversations arising from offered access to assistants; LMSYS-Chat-1M documents conversations on public chat platforms. These resources make varied uses observable, but their users and interfaces have specific selection mechanisms \citep{zhao2024wildchat,zheng2024lmsys}. A support log, an internal search, and a conversation with a general-purpose assistant can all come from real people while describing different populations and objectives.

The working paper \emph{How People Use ChatGPT} also illustrates the importance of the unit and exclusions. Its protocol distinguishes several samples and weighting schemes; its analysis of consumer usage does not indiscriminately cover all accounts, messages, and products. Privileged access to platform data improves observation of a defined population; it does not eliminate the need to specify its boundaries \citep{chatterji2025use}.

At the other extreme, a synthetic corpus can provide controlled coverage and reveal rare cases. The PersonaGen-1M preprint proposes a large bank of personas and annotated requests, but its validation does not establish that frequencies correspond to a population of buyers. Volume and internal diversity document the produced resource; they do not suffice to validate a market estimate \citep{personagen2026}. The problem is therefore not a distinction between “real” and “fake” prompts. It is the fit between the corpus production mechanism and the claimed inference.

Sampling theory explains why increasing the number of observations does not necessarily compensate for a flawed selection mechanism \citep{meng2018bigdata}. If a region of the target population has no chance of being collected, the positivity, or overlap, condition is violated. An empty cell in a finite sample does not by itself establish this structural impossibility. In either case, without observations of non-English requests or additional assumptions, reweighting English requests cannot estimate behavior in other languages.

\subsection{The Need Is Not Always Specified in the First Turn}
For GEO visibility, a source can be absent from the first answer and cited after a follow-up. A first-answer citation rate, a rate per turn, and the probability of at least one citation during a conversation therefore have different units and denominators. Replacing a trajectory with one exhaustive prompt changes the answer opportunities being measured.

Search Arena provides a setting for multilingual, multi-turn search conversations. The design nevertheless encourages requests requiring web search and relies on volunteer participants \citep{miroyan2026searcharena}. Laban et al. compare instructions provided all at once with their gradual disclosure in simulated conversations; the observed differences show that these presentation modes are not interchangeable for the tasks studied \citep{laban2026multiturn}. Neither study directly supplies a distribution of purchasing journeys.

An evaluator who turns a hesitant conversation into an exhaustive initial question may remove part of the assistant's work: clarifying, requesting a constraint, or correcting an interpretation. The evaluation then measures the resolution of an already specified need. This choice can be appropriate for isolating a capability; it must be declared.

To evaluate a complete interaction, $K$ must be extended to a user policy $\Omega$ that produces the next message as a function of the history. The trajectory and its length can depend on the system being tested. Replaying identical messages after different answers defines a controlled test but can create incoherent exchanges; allowing a simulator to react evaluates the assistant–simulator pair. A comparison of utility then requires validation of the simulator's behavior and a stopping rule, not merely a collection of initial prompts.

\subsection{Language, Location, and Assumed Knowledge}
Language and location must be distinguished. As an exploratory illustration, a preprint by Żatuchin crosses these factors on a generative search interface and an API. Its 234 usable runs show differences in named providers under some conditions; the control in another category does not reproduce the same effect. The small number of needs and short collection window rule out treating this as a universal law of the linguistic market \citep{zatuchin2026language}.

Regardless of the evidential strength of this pilot, a corpus construction principle follows: translating a prompt does not ensure preservation of all relevant conditions. Local availability, professional vocabulary, regulation, and users' familiarity with solutions can change the problem. A bank of requests generated from expert descriptions may also assume knowledge of the very terms users are trying to discover. Equivalence must be assessed at the level of the need and available knowledge, not textual fluency alone.

\section{Generative Visibility: Intervention, Competition, and Denominators}
\subsection{From Supplied Context to Endogenous Retrieval}
Aggarwal et al.'s GEO benchmark studies the optimization of content representation in answers with sources supplied to the generator. SAGEO Arena reevaluates retrieval, reranking, and generation after modifying a document, while initially selecting a target already among the documents admitted to generation \citep{aggarwal2024geo,kim2026sageo}. These protocols concern different interventions. A gain obtained in the former does not suffice to predict the result of the latter, still less the organic exposure of a newly published page.

The distinction connects to the historical problem of items made observable by a retrieval system: a document absent from the context ordinarily cannot be cited from that context. Yet absence of citation, absence of retrieval, and absence of contribution are not equivalent. Copies, parametric memory, and substitute sources can also play a role. Discovery, retrieval, selection, contribution to content, displayed citation, and brand mention must be measured separately. Clicks, conversions, and human effects require other observations.

Attribution research already distinguishes source support from the faithfulness of the attribution. AIS tests whether contextualized information is supported by a supplied source \citep{rashkin2023ais}. ALCE evaluates statement coverage by citations and the relevance of those citations. Its citation-precision test can remove a citation to assess the support provided by the remaining citations, while holding the answer fixed \citep{gao2023alce}. Wallat et al. explicitly define causal citation faithfulness and examine it through document perturbations and answer regeneration; their comparisons condition on recovering the original claim \citep{wallat2025faithfulness}. These studies therefore include intervention-based precedents. The contrast below specifies the source intervention, context policy, and substantive answer score for visibility measurement under competition; it does not claim to introduce causal attribution evaluation. Documentary support for an already generated claim and a change in the generated answer remain distinct targets.

\subsection{An Explicit Source Counterfactual}
Fix a prompt $p$, conditions $c$, and an ordered list of retrieved sources $D$. For a source $d$, a rule $\Phi_0(D,d)$ removes that source and preserves the order of the others, without new retrieval; $\Phi_1(D,d)=D$. The rule must specify whether freed slots are removed, neutralized, or filled, together with the context budget and treatment of copies. It must also declare whether removing a source compresses the context and shifts later sources, or preserves slots of documented length. Filler content is part of the intervention and is not presumed neutral. Liu et al. show position-dependent accuracy in multi-document question answering, including poorer performance when relevant information is in the middle of the context \citep{liu2024lostmiddle}. This motivates documenting position; it establishes neither the neutrality of padding nor a GEO visibility effect.

With a conditional generator $G$ and a substantive function $h$, the local effect is:

\begin{equation}
\begin{aligned} \tau_d(p,c,D)&=\mathbb E[h(Y_1)-h(Y_0)],\\ Y_a&\sim G(\cdot\mid p,c,\Phi_a(D,d)). \end{aligned}
\end{equation}
The outputs are realizations from the two experimental distributions; the expectation permits comparison of stochastic generations without assuming two deterministic texts. $h$ can score the presence of a predefined argument, a recommendation, or factual fidelity. The indicator “cites $d$” alone is insufficient: removing a source identifier can mechanically remove its citation without changing the answer's information. The judge and its instruction must remain identical in both arms.

This contrast can be estimated through repeated generations in both conditions, pairing the contexts; the number of independent contexts governs the scope beyond a single case. Removing an entire documentary lineage targets an intervention different from removing an isolated URL. Redundancy and complementarity imply that \emph{leave-one-out} effects do not necessarily add up. A Shapley-type allocation would require defining and evaluating other document coalitions; it is not equated here with the removal effect.

For an intervention in production, $D$ must no longer be fixed. Let $\mathbf z=(z_b,\mathbf z_{-b})$ be the complete vector of source interventions: $z_b\in\{0,1\}$ represents the focal publisher's intervention and $\mathbf z_{-b}$ those of the other publishers. The relevant output is $Y_b(\mathbf z)$. We write $m_b(z_b,\mathbf z_{-b})$ for $m_b((z_b,\mathbf z_{-b}))$. Its average direct effect under a competitor policy $\nu$ is:

\begin{equation}
\begin{aligned} m_b(\mathbf z)&=\mathbb E[h_b(Y_b(\mathbf z))],\\ \tau_b(\nu)&=\mathbb E_{\mathbf Z_{-b}\sim\nu}\bigl[m_b(1,\mathbf Z_{-b})\\&\qquad\qquad-m_b(0,\mathbf Z_{-b})\bigr]. \end{aligned}
\end{equation}
Retrieval and reranking are then executed in each condition. An outcome for one publisher that depends on others' treatments constitutes \textbf{interference} in the sense of causal inference \citep{hudgens2008interference}. The formula defines a target; it does not ensure identification from observational traces. Identification would require an appropriate experimental design or additional assumptions about assignment and interactions. Comparing two dates without recording the documentary state, competing interventions, and system versions confounds these changes. A comparison with fixed prompts does not remove this interference.

The corresponding \textbf{indirect effect} holds publisher $b$'s intervention at zero and changes the competitors' policy from $\nu_0$ to $\nu_1$:

\begin{equation}
\begin{aligned} \iota_b(\nu_1,\nu_0;0)&=\mathbb E_{\mathbf Z_{-b}\sim\nu_1}[m_b(0,\mathbf Z_{-b})]\\ &\quad-\mathbb E_{\mathbf Z_{-b}\sim\nu_0}[m_b(0,\mathbf Z_{-b})]. \end{aligned}
\end{equation}
This is a spillover effect under interference, not a mediation effect. It formalizes the publisher's question: can my visibility change while my own content remains unchanged? Prompt distribution, engine settings, scoring, and the initial documentary state must be held comparable; retrieval is rerun under each policy. A negative value would describe a loss attributable to that policy change under the specified design. An observed decline between dates does not, by itself, identify this effect.

Under a common protocol and initial conditions, the \textbf{joint total effect} compares the regimes $(z_b=0,\nu_0)$ and $(z_b=1,\nu_1)$:

\begin{equation}
\begin{aligned} \tau_b^{\mathrm{tot}}(\nu_1,\nu_0)&=\mathbb E_{\mathbf Z_{-b}\sim\nu_1}[m_b(1,\mathbf Z_{-b})]\\ &\quad-\mathbb E_{\mathbf Z_{-b}\sim\nu_0}[m_b(0,\mathbf Z_{-b})]\\ &=\tau_b(\nu_1)+\iota_b(\nu_1,\nu_0;0). \end{aligned}
\end{equation}
This identity allows interactions but depends on the reference policy. A publisher observes a temporal score change, not automatically this causal total effect. At least four mechanisms can contribute, apart from changes in formulation and scoring: its own intervention, competing interventions, engine drift, and a change in corpus composition. Equation (5) separates response change from composition change descriptively. With suitable controls or assignment, Equations (7) and (8) target the direct and competitive spillover effects. Neither a time comparison nor the total-effect identity uniquely allocates the observed change among these mechanisms: that requires additional interventions, assumptions, and a reference order.

\subsection{Recognition and Discovery: Bounded Illustrations}
Sharma's preprint compares, in December 2025, three named prompts and seven discovery prompts for each of 112 products through two APIs. It reports, for GPT-4o-mini, 334 successes out of 336 named trials versus 26 out of 784 discovery trials; for Sonar, 317 out of 336 versus 65 out of 784 \citep{sharma2026discovery}. Success means the exact presence of the name. The novelty criteria in the open-ended requests and the naming of products in the other condition prevent interpreting the difference as a pure paraphrase effect.

We use this study as an exploratory illustration of different tasks, not as a market estimate. The same caution applies to Żatuchin's 234 runs and the synthetic PersonaGen corpus: they provide cases to examine, not independent replications establishing a language-by-market law or human frequencies. Both share part of their provenance. The status, scale, and scope of these sources are presented in Table 3; the identification arguments in the preceding sections do not depend on them.

\subsection{What the Compared Evidence Supports}
\begin{table*}[t]
\centering
\caption{Scale, status, and inferential boundaries of visibility studies.}
\small
\renewcommand{\arraystretch}{1.12}
\begin{tabular}{@{}>{\raggedright\arraybackslash}p{\dimexpr 0.3\linewidth-1.3333333333333333\tabcolsep\relax}>{\raggedright\arraybackslash}p{\dimexpr 0.3\linewidth-1.3333333333333333\tabcolsep\relax}>{\raggedright\arraybackslash}p{\dimexpr 0.4\linewidth-1.3333333333333333\tabcolsep\relax}@{}}
\toprule
\textbf{Study, status, and period} & \textbf{Scale examined} & \textbf{Contribution and inferential boundary} \\ \midrule
GEO; KDD 2024; version of 28 June 2024 & 10,000 queries, including 1,000 test queries & Supplied sources: conditional effect, not organic discovery \\
SAGEO; KDD 2026; version of 7 August 2026 & 2,700 test queries from nine datasets; 171,003 documents & Pipeline rerun; controlled distribution and initially retrieved target \\
Grossman; SIGIR 2026; 7–8 Dec. 2025 & 11,500 queries in the main collection & Mobile, Newark, two days; activation and sources on a benchmark, not a representative market \\
Sharma; 2026 preprint; Dec. 2025 & 112 products; 336 + 784 trials per API & Exact mention, nonequivalent conditions; illustration \\
Xu; 2026 preprint; Mar.–Apr. 2026 & 55,393 trending queries & Activation by observed categories, not a causal wording effect \\
Żatuchin; 2026 preprint; 29–30 Aug. 2026 & 234 usable runs, 12 rejected; few needs & Exploratory comparison, partial design, and nonconvergent control \\
PersonaGen; preprint of 30 August 2026 & 1,031,732 synthetic personas; public subset of 14,955 personas & Construction and internal validation, not estimation of human frequencies \\
\bottomrule
\end{tabular}
\end{table*}
The comparison first establishes that these studies do not observe the same variable under different brands: they change the unit, context, activation, and success criterion. Second, experiments with supplied contexts and interface audits do not answer the same causal question. Finally, very large corpus sizes compensate for neither nonequivalent tasks nor missing overlap. The rightmost column defines what each result can support; it prevents counting heterogeneous designs as repeated confirmations of the same effect.

\subsection{A Secondary Demonstrator Using Published Aggregates}
In this subsection, the textual subscripts act, cond, and req designate probabilities; $p$ and $p_{ij}$ elsewhere denote prompts.

Grossman et al. report 65.6\% AI Overviews activation across 11,500 queries for their main collection through SerpAPI on 7–8 December 2025. The setting is mobile, in Newark, New Jersey. ORCAS, Amazon Retail, and ELI5 have activation rates of 51.5\%, 17.4\%, and 94.6\%, respectively. Table 4 separates these published inputs from our calculations \citep{grossman2026disrupts}. The source percentages are rounded. The numerical intervals in the table are plug-in bounds on these panel frequencies, not guaranteed population bounds; additional decimal places do not imply greater empirical precision.

Let $A_{\mathrm{act}}$ denote activation and $M$ a mention within the module. If $p_{\mathrm{act}}=\Pr(A_{\mathrm{act}})>0$ and $p_{\mathrm{cond}}=\Pr(M\mid A_{\mathrm{act}})$, presence per query is $p_{\mathrm{req}}=p_{\mathrm{act}}p_{\mathrm{cond}}$. This is a different quantity from $p_{\mathrm{cond}}$. Here, $M$ concerns the generative module exclusively, not the entire results page.

\begin{table*}[t]
\centering
\caption{Published activation and derived calculations; the final column is hypothetical. The total contains the subsets and is not an additional stratum.}
\small
\renewcommand{\arraystretch}{1.12}
\begin{tabular}{@{}>{\raggedright\arraybackslash}p{\dimexpr 0.2\linewidth-1.6\tabcolsep\relax}>{\raggedright\arraybackslash}p{\dimexpr 0.12\linewidth-1.6\tabcolsep\relax}>{\raggedright\arraybackslash}p{\dimexpr 0.15\linewidth-1.6\tabcolsep\relax}>{\raggedright\arraybackslash}p{\dimexpr 0.28\linewidth-1.6\tabcolsep\relax}>{\raggedright\arraybackslash}p{\dimexpr 0.25\linewidth-1.6\tabcolsep\relax}@{}}
\toprule
\textbf{Published set} & \textbf{Queries} & \textbf{Activation $p_{\mathrm{act}}$} & \textbf{Bound on $p_{\mathrm{req}}$, without mention data} & \textbf{$p_{\mathrm{req}}$ if $p_{\mathrm{cond}}=40\%$ (assumption)} \\ \midrule
Entire benchmark & 11,500 & 65.6\% & [0; 65.6\%] & 26.24\% \\
\midrule
ORCAS & 5,000 & 51.5\% & [0; 51.5\%] & 20.60\% \\
Amazon Retail & 500 & 17.4\% & [0; 17.4\%] & 6.96\% \\
ELI5 & 1,000 & 94.6\% & [0; 94.6\%] & 37.84\% \\
\bottomrule
\end{tabular}
\end{table*}
The 65.6\% rate describes the whole corpus, which includes ORCAS: it must not be mixed with ORCAS as though they were disjoint strata. Our three-subcorpus panel deliberately includes Amazon Retail and ELI5, the minimum and maximum activation rates in the published comparison, with ORCAS as a third setting. Selecting these strata is part of specifying the support of $\Pi$, not a neutral feature of the input data. It makes a contrast visible; it does not estimate the typical sensitivity of another corpus. Within this selected support, weights $(0.2,0.6,0.2)$ and $(0.2,0.2,0.6)$ yield \textbf{39.66\%} and \textbf{70.54\%} activation, a difference of \textbf{30.88 percentage points}. This contrast is attributable to the weights conditional on the chosen strata; its magnitude also depends on that selection.

A reference retaining the collected counts within this restricted panel uses weights $(5000,500,1000)/6500$, approximately $(0.769,0.077,0.154)$. It yields \textbf{55.5\%} activation, not the full benchmark's 65.6\%. This reference score lies numerically between the two scenarios. Its weight vector, however, is outside the following class, which fixes ORCAS at 20\%.

More generally, retain 20\% for ORCAS and allocate $\beta\in[0.2;0.6]$ to Amazon, with the remainder assigned to ELI5. Then:

\begin{equation}
\begin{aligned}p_{\mathrm{act}}^{\mathrm{panel}}(\beta)&=0.2(0.515)+\beta(0.174)\\&\quad+(0.8-\beta)(0.946)\\&=0.8598-0.772\beta.\end{aligned}
\end{equation}
The envelope of this entire class is $[0.3966;0.7054]$, computable from its endpoints. A class with no further constraints on these three strata yields $[0.174;0.946]$. These are \textbf{sensitivity envelopes across normative estimands}, evaluated using the published rates, not identified sets for an unknown usage population. Their difference expresses the strength of the weighting restrictions; neither is a confidence interval.

Activations alone cannot determine $p_{\mathrm{cond}}$. They provide the bounds $0\leq p_{\mathrm{req}}\leq p_{\mathrm{act}}$. The final column therefore uses \textbf{an illustrative assumption}: a conditional mention rate of 40\% in every set. Only under this assumption do our two panels yield 15.864\% and 28.216\% presence per query, despite the same conditional rate of 40\%. For $p_{\mathrm{req}}>0$, replacing the denominator “all queries” with “activated modules” multiplies the rate by $1/p_{\mathrm{act}}$, approximately 2.52 or 1.42 in the two panels. No observed brand ranking is inferred from this calculation.

The same article distinguishes Natural Questions from its keyword transformation, with activation rates of 86.2\% and 76.5\% for 1,000 queries in each set. A mixture policy assigning weight $\gamma$ to the first form yields $p_{\mathrm{act}}^{\mathrm{NQ}}(\gamma)=0.765+0.097\gamma$, or 81.35\% at equal weights. This is a second axis of sensitivity on this particular support. The aggregates do not permit crossing it with the three preceding strata to construct a complete factorial design. Differences in formulation, selection, and denominator must not be merged in the absence of common cells.

The residual reconstruction requires the five named categories to be disjoint subsets of the 11,500 records. We adopt the nine-category partition described in Section 3 of Grossman et al. and the exact denominators in their Table 1; the aggregate rates alone would not establish that partition. On this basis, the five categories contain 8,500 query records, leaving 3,000. The unit reconstructed is a query record, not a need: NQ and its keyword transformation share underlying questions, and independence of needs is not assumed.

Rounding is amplified by this subtraction. Assuming each published rate is a binary-event proportion rounded to the nearest 0.1 percentage point, compatible integer activation counts lie in $[7539;7549]$ for the total and $[5233;5237]$ for the five categories. The remainder is therefore $[2302;2316]/3000$, reported with outward rounding as \textbf{[76.7\%; 77.2\%]}. Nearest rounding is an explicit assumption, not a documented source convention. Propagating the same rounding errors without integer constraints gives the wider envelope \textbf{[76.6\%; 77.3\%]}; its half-width is about 0.33 percentage point, versus 0.05 for the count-weighted three-category reference. These are arithmetic compatibility envelopes, not confidence intervals or recovered exact counts. Separate residual category rates and individual responses remain unavailable from these inputs, precluding arbitrary full-benchmark reweighting and an observed crossing of formulations and strata.

Finally, Xu et al. report 13.7\% overall activation, 64.7\% for queries classified as questions, and 9.5\% for the others, across 55,393 US trending queries collected from 13 March to 21 April 2026 \citep{xu2026overviews}. These observed categories are not a random assignment of reformulations. The difference from Grossman therefore measures neither an isolated temporal change nor a causal effect of syntax. The demonstrator establishes a consequence of conventions on published aggregates; a visibility measure still requires joint observations of mentions and activation.

\section{Uncertainty and a Computable Protocol}
\subsection{Weights, Dependencies, and Efficient Evaluation}
To illustrate the distinction between repetition, reformulation, and sampling, consider $Z_{ij\ell}=\mu+a_i+\zeta_{ij}+e_{ij\ell}$, with independent centered components and variances $\sigma_U^2$, $\sigma_P^2$, and $\sigma_E^2$. For fixed normalized weights, the variance of the hierarchical estimator is:

\begin{equation}
\begin{aligned} \mathrm{Var}(\widehat\theta)&=\sigma_U^2\sum_i w_i^2\\ &\quad+\sigma_P^2\sum_i w_i^2\sum_jv_{ij}^2\\ &\quad+\sigma_E^2\sum_i w_i^2\sum_j\frac{v_{ij}^2}{r_{ij}}. \end{aligned}
\end{equation}
With constant $n,k,r$ and equal weights, this reduces to $\sigma_U^2/n+\sigma_P^2/(nk)+\sigma_E^2/(nkr)$. With uniform variant weights and unequal $w_i$, weight concentration enters through $\sum_iw_i^2$; under this homogeneous model, the relative effect is $n\sum_iw_i^2$. This is not a universal formula for the design effect. Dependencies between sessions, temporal shocks, heteroskedasticity, and weight estimation require additional terms or models.

For an entirely fixed panel, variation among its needs is initially descriptive heterogeneity; it becomes sampling error only relative to an explicit population or model. Treating paraphrases as independent situations creates pseudoreplication. Work on multiple generations and moment methods helps estimate some components without resolving missing coverage \citep{zhang2026multiple,lior2025reliableeval}.

Efficient evaluation then raises a conditional question: how can the score of a defined benchmark be estimated with fewer observations? Lalor et al. draw on item response theory; tinyBenchmarks and PromptEval exploit performance structure to reduce the required evaluations \citep{lalor2016irt,maiapolo2024tinybenchmarks,polo2024prompteval}. These savings depend on assumptions about generalization across items, models, or prompts. They do not identify the distribution of human needs. Estimation efficiency on a benchmark and external validity of its target must therefore be assessed separately.

\subsection{A Protocol with Verifiable Outputs}
Table 5 links each implementation step to the four decisions in the conclusion: \textbf{D1}, specify the protocol; \textbf{D2}, justify support and weights; \textbf{D3}, define source and competitor interventions; \textbf{D4}, compare under common weights. One step may serve several decisions.

Together, these steps make the chain of inference inspectable: each score is linked to a declared target, observable states, a scoring rule, and admissible weights. Retaining records at these levels allows changes in execution, coding, or composition to be examined before interpreting a difference between systems or dates. This joint specification supports reproducible calculation and scrutiny of its assumptions; representativeness and causal identification still require the conditions stated in the preceding sections.

\begin{table*}[t]
\centering
\caption{Seven implementation steps linked to the four decisions.}
\small
\renewcommand{\arraystretch}{1.12}
\begin{tabular}{@{}>{\raggedright\arraybackslash}p{\dimexpr 0.08\linewidth-1.3333333333333333\tabcolsep\relax}>{\raggedright\arraybackslash}p{\dimexpr 0.78\linewidth-1.3333333333333333\tabcolsep\relax}>{\raggedright\arraybackslash}p{\dimexpr 0.14\linewidth-1.3333333333333333\tabcolsep\relax}@{}}
\toprule
\textbf{Step} & \textbf{Required action and record} & \textbf{Decision(s)} \\ \midrule
1 & \textbf{Define the target and intervention.} Specify the construct, unit, population or normative priority, period, metric, and denominator. Define the boundary of $S$ and the inputs of $H$ before choosing controls. & D1, D3 \\
2 & \textbf{Document annotation and overlap.} Retain accessible events or vignettes, the rules $A_\eta$, disagreements, exclusions, and links between families. Identify missing cells; do not assign them an observed performance. & D1, D2 \\
3 & \textbf{Fix or bound the weights.} Declare the origins of $w$ and $v$, constraints on $\Pi$, and reasons for each scenario. Distinguish estimated empirical weights, conventions, and uncertainty classes; reserve control variants for final checks. & D2 \\
4 & \textbf{Trace execution.} Record the complete prompt, accessible system instructions, history, parameters, date and time, interface, model/version identifier when exposed, and request and response fingerprints. A stable identifier does not guarantee the absence of silent updates: interleave conditions, use sentinel controls, and segment collection if a break is suspected. & D1, D3 \\
5 & \textbf{Code outputs and validate the judge.} Apply the state scheme in Table 6 and the attribution coding in Section 8.2, drawing on Section 7.1. Retain the raw text and coding rule; audit outputs with independent judgment when necessary. Also version the judge's instruction and check scoring differences on constant outputs. & D1 \\
6 & \textbf{Calculate sets and their uncertainty.} Produce cell-level scores, the score under the main convention, and extrema under $\Pi$. Compare systems under the same weights. Separately add appropriate sampling or repetition uncertainty; publish the constraints active at the extrema. Multiverse analysis provides a transparency precedent for defensible analytical choices \citep{steegen2016multiverse}. & D2, D4 \\
7 & \textbf{Make the comparison inspectable.} Publish configurations, shareable data, calculations, corpus version, and differences from the previous collection. A source-removal contrast must record sources, copies, order, budget, substitution, and the relevant state of competitors. & D3, D4 \\
\bottomrule
\end{tabular}
\end{table*}
\begin{table*}[t]
\centering
\caption{Coding scheme for observed states.}
\small
\renewcommand{\arraystretch}{1.12}
\begin{tabular}{@{}>{\raggedright\arraybackslash}p{\dimexpr 0.25\linewidth-1.3333333333333333\tabcolsep\relax}>{\raggedright\arraybackslash}p{\dimexpr 0.35\linewidth-1.3333333333333333\tabcolsep\relax}>{\raggedright\arraybackslash}p{\dimexpr 0.4\linewidth-1.3333333333333333\tabcolsep\relax}@{}}
\toprule
\textbf{Observed state} & \textbf{Minimal coding} & \textbf{Consequence for the denominator} \\ \midrule
Activated module, usable answer & Activation = 1; separate mention and citation fields; support if assessable & Included in per-query rates and among activations \\
Valid non-activation & Activation = 0; presence in module = 0 & Included per query; absent from the conditional rate \\
Explicit refusal & Refusal status; activation coded separately & Retain, then apply the target convention \\
Empty output received & Empty status; known or undetermined cause & No automatic recoding as non-activation \\
Missing collection or technical error & Missing status; reason and attempt & Neither observed brand absence nor silent exclusion \\
\bottomrule
\end{tabular}
\end{table*}
For each usable answer, retain separate fields for retrieval and context selection when observable, displayed citations, and brand mentions; link each evaluated claim to source identifiers and a support judgment. When attribution is evaluated, distinguish AIS-style documentary support, statement coverage by citations, and citation precision as described in Section 7.1; declare the units, denominators, sources accessible to the judge, and conventions for answers without citations. An observed citation with an inaccessible source has unassessable support, not automatically zero support; unavailable retrieval or selection traces likewise remain missing. These measures do not establish the causal contribution defined in Section 7.2.

Refusal and emptiness can be system behaviors or collection problems: available traces determine what can be classified. For a population of assigned queries, deleting missing collections generally estimates a quantity conditional on observation. If the missingness rate is known and the score bounded, worst-case bounds are preferable to unjustified zero imputation. Fingerprints support traceability; they reveal neither model weights nor all internal transformations.

\subsection{An Extension of Datasheets, Not a Competing Checklist}
Datasheets already organize documentation of motivation, composition, collection, processing, uses, distribution, and maintenance \citep{gebru2021datasheets}. Our card adds precise links to the estimated quantity and the intervention. Table 7 describes the methodological additions, not a claim to have invented documentation.

\begin{table*}[t]
\centering
\caption{Extending Datasheets to document estimands and interventions.}
\small
\renewcommand{\arraystretch}{1.12}
\begin{tabular}{@{}>{\raggedright\arraybackslash}p{\dimexpr 0.26\linewidth-1.3333333333333333\tabcolsep\relax}>{\raggedright\arraybackslash}p{\dimexpr 0.38\linewidth-1.3333333333333333\tabcolsep\relax}>{\raggedright\arraybackslash}p{\dimexpr 0.36\linewidth-1.3333333333333333\tabcolsep\relax}@{}}
\toprule
\textbf{Existing documentation field} & \textbf{Addition for a prompt corpus} & \textbf{Expected output} \\ \midrule
Motivation and uses & Estimand, unit, denominator, intervention & Target specification and defined contrast \\
Composition and collection & Families, variants, $A_\eta$, overlap & Annotation scheme and missing cells \\
Processing and scoring & Origins of $w,v$, class $\Pi$, judge instruction & Weights, constraints, and cell-level scores \\
Maintenance and distribution & Engine state, fingerprints, competition, and changes & Version and differences between collections \\
\bottomrule
\end{tabular}
\end{table*}
A fixed core permits comparison at a constant corpus; a refreshed panel tracks contemporary demand; an experimental component explores new situations. Results must remain separate. The core neutralizes neither documentary change nor personalization nor updates: it fixes one component of the instrument. The appendix shows how the same requirement applies to the article's bibliographic corpus.

\par\ifdim\dimexpr\pagegoal-\pagetotal\relax<8\baselineskip\relax\newpage\fi
\section{Discussion and Limitations}
\subsection{Why Prompts Belong to the Measurement Instrument}
One objection is that every average depends on the cases selected. This is true, but does not fully explain the role of prompts. With a predefined population, another sample may estimate the same quantity differently. For an evaluation panel, one must also specify whether the situations and weights represent an external population or define the score's answer market. This determines what can be concluded from a corpus change.

A prompt also acts as an instruction. Open-ended recommendation requests and comparisons naming two brands create different opportunities for presence: they assign different tasks, rather than merely rephrase one task. With document retrieval, instructions can also alter search queries, retrieved documents, and which sources compete. Prompts thus help create the conditions under which visibility is observed.

Two protocols can thus address different questions. Supplying the generator with the same documents allows its answers to be studied within a fixed context. Letting the engine retrieve sources for each prompt evaluates a pipeline that includes document selection. Results from the first protocol do not automatically transfer to the second: effects on answers given supplied documents must be distinguished from effects on visibility in the full engine.

Scoring adds another dependence. When an LLM judge is used, changing its instruction can change the evaluation of an unchanged answer. Changes in answers must therefore be separated from changes in their conversion into scores. Situation selection, answer generation, and scoring require separate documentation, rather than an undifferentiated label of “prompt sensitivity.”

Each comparison should thus be interpretable within a declared scope. HELM proposes explicit coverage and multiple evaluation dimensions; prompt-conditional rankings likewise state a target dependent on the requests considered \citep{liang2023helm,frick2025p2l}. Here, a ranking is robust if it holds throughout the selected admissible class, with the same weights applied to all systems. Otherwise, the conclusion must specify which choices determine the ordering. An unstable ranking proves neither system equality nor an inability to choose for a particular use: the comparison must be tied more precisely to that use.

\subsection{What the Developments Establish and What Remains Open}
The fictional example proves a conditional possibility; extrema over finite or convex classes follow from the model; the demonstrator establishes the arithmetic magnitude of a weighting change on published aggregates. None of these results estimates a prompt's effect on a human population. The claimed novelty lies in connecting problems and operationalizing them for generative visibility measurement, not in inventing partial identification, weighted variance, or the counterfactual.

Empirical validation should compare construction procedures against an observed population, test the constraints chosen for $\Pi$, measure the fidelity of annotations and judges, and then examine whether transported conclusions predict observations outside the corpus. Another experiment should compare context removal with intervention on the full pipeline, with treatment assignment, replications by context, and an explicit competition protocol. Dialogue simulator performance and the detection of silent breaks are also open questions.

\subsection{Limitations of the Evidence Base and Aggregates}
The survey is selective and predominantly English-language. Publication statuses and partial readings are declared; several visibility preprints are retained as illustrations of limited scope. Convergence in vocabulary across these studies is not an accumulation of independent replications. Annotating a situation does not reveal a true latent need, and a documented judge can remain inadequate.

The rates reported by Grossman et al. describe a particular collection and specific sets. Rounding, dependencies between queries, and individual-level data not reanalyzed here preclude deriving valid frequentist intervals from them. Visibility bounds without mentions are wide precisely because information is missing. Without joint data on variants and strata, we do not present their Cartesian product as an observed multiverse. A poorly justified class $\Pi$ can make a result artificially robust or undetermined.

Finally, neither exposure to the module nor mention nor source attribution directly measures attention, trust, purchasing, or persuasion. These effects require human units, observations, and their own designs. The proposed instrument clarifies what is missing to study them; it does not identify them by proxy.

\section{Conclusion}
A GEO visibility score summarizes the observed presence of sources or brands in answers produced under a given protocol. The prompt corpus and its weights delimit the answer market over which this visibility is evaluated. Prompts play a dual role: they select the situations covered and elicit the system's behavior, potentially altering retrieval and the sources placed in competition. When an LLM judge is used, its instruction adds another dependence on formulation. This connection gives meaning to the article's central distinction: an influence on the output is a behavioral fact; participation in the definition of measurement is a property of the evaluation design.

The developments presented here clarify the consequences of this distinction. The analytical example establishes that changing weights can reverse a ranking when system profiles cross between strata. The demonstrator based on published aggregates shows how panel composition and denominator conventions change the quantities reported. It concerns the activation of generative answers; it is not an observed measure of a publisher's visibility. These results explain why a difference between two scores must be examined in light of the conditions under which each was constructed.

The methodological contribution organizes this examination around four decisions.

\textbf{D1. Make explicit what the score measures and how it is produced.} Defining the target, observed units, annotation rules, formulations, execution environment, and scoring links the result to a precise protocol. The chosen denominator, the treatment of missing observations, and, where applicable, the judge and its instruction are part of this definition. This documentation allows the scope of the score to be assessed and its calculation to be reproduced.

\textbf{D2. Justify the situations covered and the weights assigned to them.} When several weighting schemes remain admissible, a single score conceals some of the choices on which the result depends. A declared and justified class allows the values compatible with these choices to be calculated. Their interpretation must remain explicit: a sensitivity envelope across measurement conventions, an identified set derived from data and assumptions about a population, and sampling uncertainty address different questions.

\textbf{D3. Define the intervention before concluding that a source makes a causal contribution.} The presence of a citation, documentary support for a statement, and a source's effect on an answer are distinct objects. The proposed removal contrast specifies how to study the last of these in a controlled context. Extending it to a full engine requires accounting for retrieval, reranking, and competitor interventions. A temporal change in visibility alone cannot disentangle these mechanisms.

\textbf{D4. Compare on a common basis and make the result's dependencies visible.} Comparing systems under the same weights avoids attributing differences caused by corpus composition to system behavior. When weights vary within an admissible class, the contrast must be calculated under each common weighting scheme, rather than by freely subtracting score envelopes. Temporal monitoring also requires examining changes in the engine, formulation, scoring, and competition. A ranking can then be presented with the conditions under which it remains robust.

This framework provides computable objects and rules for interpretation; its empirical usefulness remains to be tested. The survey is selective, and neither the analytical examples nor the recalculations of aggregates constitute validation on a population of users or a demonstration of the effectiveness of GEO optimization. Future research should compare corpora against observed usage, evaluate annotations and judges, and then test the conclusions in other situations and in suitable experimental designs.

\par\ifdim\dimexpr\pagegoal-\pagetotal\relax<6\baselineskip\relax\newpage\fi
The value of a GEO score thus lies in being able to specify which answer opportunities it describes, under what conditions it can be compared, and what its changes actually allow us to conclude. Documenting these conditions is an integral part of visibility measurement.

\onecolumn
\section*{AI Assistance Statement}
This manuscript was prepared with assistance from Codex (OpenAI) for writing support, translation, and file preparation. This assistance does not constitute independent validation of the sources or arguments. Validation of the sources and arguments, the selection and use of references, and responsibility for publication rest with the human author of this study.

\section*{Availability and Reproducibility}
The manuscript sources, bibliography, reading register, transcribed aggregates, and calculation programs accompany this version. No new experimental collection from users or generative systems is reported. The demonstrator is a secondary recalculation of published aggregates, with scenarios and assumptions separated from data; the ranking example is fictional. Private preparatory documents are not distributed. Rights attached to sources must be preserved in any redistribution.

\appendix
\section{Corpus Card for This Survey's Documentary Corpus}
This card applies the article's requirements to its own synthesis instrument. The unit is the scientific work, grouping its versions. There is neither a numerical score per article nor an estimator of the prevalence of findings: bibliographic population weights would therefore be irrelevant here. Coverage choices nevertheless influence the arguments available.

\begin{table}[!ht]
\centering
\caption{Corpus card for this survey’s documentary selection.}
\small
\renewcommand{\arraystretch}{1.12}
\begin{tabular}{@{}>{\raggedright\arraybackslash}p{\dimexpr 0.25\linewidth-1.0\tabcolsep\relax}>{\raggedright\arraybackslash}p{\dimexpr 0.75\linewidth-1.0\tabcolsep\relax}@{}}
\toprule
\textbf{Dimension} & \textbf{Statement for the present article} \\ \midrule
Objective and unit & Conceptual synthesis of measurement decisions; one work, versions grouped \\
Realized corpus & 44 references; no exhaustive population of relevant works enumerated \\
Discovery & Initial dossier, titles, keywords, citation tracing, and extension during revision \\
Selection and annotation & Connection to construct, selection, formulation, execution, scoring, or transport; narrative coding without independent duplicate selection \\
Target and weights & Argumentative coverage; neither known inclusion probabilities nor an average of effects \\
Missing coverage & Non-English literatures and unidentified studies; unknown extent \\
Calculation provenance & Benchmark partition and aggregates reported by Grossman et al. checked in Section 3, Table 1, and Section 4.1; explicitly added scenarios \\
Shareable record & Reference-level reading register, stable sources, data, and calculations; private documents excluded \\
\bottomrule
\end{tabular}
\end{table}
Reading limitations requiring immediate notice appear in Table 9. For the other references, the register specifies the sections, tables, or methods consulted; “accessible text” does not mean “all experiments audited.”

\begin{table}[!ht]
\centering
\caption{Restricted readings and corresponding uses.}
\small
\renewcommand{\arraystretch}{1.12}
\begin{tabular}{@{}>{\raggedright\arraybackslash}p{\dimexpr 0.22\linewidth-1.3333333333333333\tabcolsep\relax}>{\raggedright\arraybackslash}p{\dimexpr 0.38\linewidth-1.3333333333333333\tabcolsep\relax}>{\raggedright\arraybackslash}p{\dimexpr 0.4\linewidth-1.3333333333333333\tabcolsep\relax}@{}}
\toprule
\textbf{Reference} & \textbf{Level actually used} & \textbf{Permitted use in this survey} \\ \midrule
Messick (1994) & ETS record and abstract & Intellectual lineage of validity; not the main methodological anchor \\
Frick et al. (2025) & PMLR record and abstract & Stated objective of prompt-conditional ranking \\
Zobel (1998) & Introduction, p. 307, indexed primary extraction & Pooling problem and stated general conclusion; no quantified effect \\
Manski (2003) & Introduction, pp. 1–2, publisher preview & Positioning of partial identification; no theorem from the book attributed without reading \\
\bottomrule
\end{tabular}
\par\smallskip\begin{minipage}{\linewidth}\footnotesize
\textbf{Reading note.} The main text of Tamer’s survey (abstract and Sections 1–5, pp. 167–192, author-hosted Review in Advance version) and all of Section 3 of Jacobs and Wallach were read in full \citep{tamer2010partial,jacobs2021measurement}. Section 2 and the reading register specify the versions, methods, and limits of these readings.
\end{minipage}
\end{table}

\clearpage
\twocolumn
\begingroup\interlinepenalty=10000
\bibliographystyle{plainnat}
\bibliography{references}
\endgroup
\end{document}